# ADOPTION OF PRECISION MEDICINE; LIMITATIONS AND CONSIDERATIONS

Nasim Sadat Mosavi and Manuel Filipe Santos

Algoritmi Research Centre, University of Minho, Guimaraes, Portugal

## ABSTRACT

*Research is ongoing all over the world for identifying the barriers and finding effective solutions to accelerate the projection of Precision Medicine (PM) in the healthcare industry. Yet there has not been a valid and practical model to tackle the several challenges that have slowed down the widespread of this clinical practice. This study aimed to highlight the major limitations and considerations for implementing Precision Medicine. The two theories Diffusion of Innovation and Socio-Technical are employed to discuss the success indicators of PM adoption. Throughout the theoretical assessment, two key theoretical gaps are identified and related findings are discussed.*

## KEYWORDS

*Precision Medicine, Adoption, Artificial Intelligence, Healthcare Big Data, Open data exchange, Genomes, Biological indicators, Standards, Internet of Things, Blockchain.*

## 1. INTRODUCTION

With the availability of healthcare big data and technological advancement, clinical practice going beyond the "one-size-fits-all" approach. Where emerging the new clinical decision-making, minimizes medical errors, cuts the cost of overtreatment, increases the quality of services offered by care providers, and saves more lives[1].

Precision medicine is an emerging approach in medical decision-making that takes into account individual genetic profile, environmental and lifestyle indicators. In 2015 the former president Obama, launched the Precision Medicine Initiative (PMI) aiming to improve the tailoring of the treatment based on individual variables[2]. PMI intends to motivate individuals to cooperate as co-researcher to manage their health by sharing their health data (e.g., genomic, genetic, longitudinal health information) throughout a trustful partnership platform, where, this cooperation results in obscuring the boundary between health and disease [3].

Projecting this approach requires early diagnosis, prevention, and tailor the treatment for the individual patient. Based on that, sharing data and linking individual patient variables to health records are expected to lead to the right drug, at the right dose to the right patient [4]. Hence, successful adoption of PM has been found in tailoring treatment to a particular patient subgroup with common molecular characteristics [5].

For the clinical application of precision medicine to be able to fit the patient with matching treatment modalities, the incorporation of various heterogeneous parameters is required. Although research is underway around the world to speed up progress on precision medicine projection, and there are research programs such as the "All of Us Research"that provide innovation opportunities to address the limitations and solutions[3], still, there has not been a clear protocol and business model for emerging the PM in healthcare.

The need for collecting individual patient data (e.g., genomes, biological indicators, demographics, administration) and integrating them into the Electronic Health Records (EHRs), identifying suitable approaches and techniques to deal with big data, transforming large, multimodal data into Machine Learning (ML) algorithms for decision-making [6],[7], dealing with data protection security and ethical issues of sharing such sensitive data, adopting standards for data exchange, empowering stakeholders with education, and proposing effective pipelines for restructuring the policies and regulations are some of the considerations for employing PM.

This study aimed to propose the major limitations and considerations for adopting PM in healthcare. The paper first explains two theories, which have been used widely by literature to address remarkable indicators for successful implementation of Information Systems (IS). After that, the main limitations and challenges for projecting PM are highlighted and finally, the paper is ended by a discussion throughout examining both theories in implementation of PM. In this assessment, we identified two considerable limitations and gaps, which are not taken into the account by both theories: DoI and Socio-tech.

## 2. THEORETICAL FOUNDATION

The aspects of adoptinginnovation/ technology have been studied for over 30 years, and literature has used multiple theories to address the success indicator for adopting new technology. The theoretical basement of this study highlights the general consideration of Information Systems (IS) adoption through the Socio-technical theory and Diffusion of Innovation (DoI). Where both theories have been widely used in scientific research from a broad variety of disciplines [8].

### 2.1. Socio-Technical Theory

According to figure 1, the socio-technical theory emphasizes that the information system contains two major interrelated subsystems: the technical and social. Whereas technical aspects address tangible factors such as technology and tasks required to convert system inputs into out-puts, social aspects mostly concentrate on organizational aspects, people and the perfect harmony between technological tools and human activities are the keys for successful adoption [9],[10].

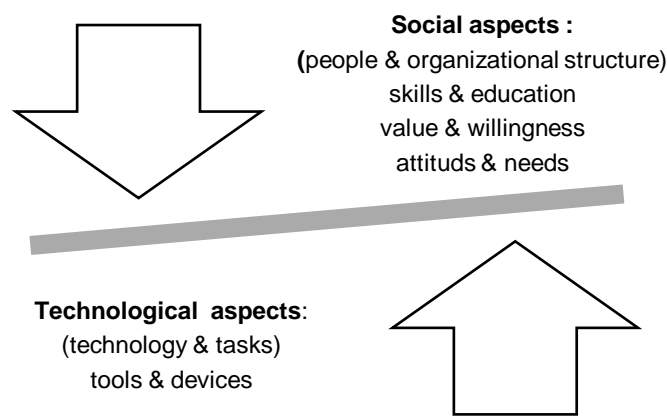

Figure 1. Socio-Tech Paradigm

## 2.2. Diffusion of Innovation Theory

From IS perspective, Diffusion of Innovation (DoI) expresses major indicators for successful adoption: compatibility ( degree in which the new technology is compatible with the existing system)**,** technical complexity(how easy/difficult is the technology to understand)**,** the relative advantage of adopting the technology (the degree to which an innovation adds value to an existing system), triability (to what extent the innovation can experiment before adoption), and observability (how the innovation provides tangible results)[9],[11].

Roger defines diffusion as the process that innovation is communicated via specific channels over time and through the member of a social platform; he used the word "technology" and "innovation" interchangeably [8]. Thus, not only technological aspects of innovation affect a successful adoption, but also, social platform and communication including, people, policies, regulations, and management influence this process.

## 3. PRECISION MEDICINE

Healthcare and medicine industries are facing dramatic changes under the influence of Artificial Intelligence (AI) techniques, Business Analytics (BA). The huge amount of data is challenging the business model of empirical medicine [1], where, answering the questions such as why does a drug work for some patients and be less effective on others? Why does medicine cause side effects on some individuals? Why do cancers influence some people and do not others? need The medical practice of tailoring each patient as a unique case.

On January 20, 2015, former President Barack Obama announced the great potential health improvement to science. The "Precision Medicine" launched by him, has been a remarkable research opportunity in a new area of medical practice for improving public health; *"Tonight, I'm launching a new Precision Medicine Initiative to bring us closer to curing diseases like cancer and diabetes and to give all of us access to the personalized information we need to keep ourselves and our families healthier."*; Hence the new approach to medical practice includes two components; a short-term which concentrates on cancer diseases and the long-term that considers other illnesses [4]. According to the U.S. National Library of Medicine, "Precision Medicine" is an emerging approach that considers individual differences such as genes, environment, and lifestyle for preventing and treating particular diseases [4], [12]. Furthermore, the institute of US National Cancer defines Precision Medicine as "a form of medicine that uses information about a person's genes, proteins, and environment to prevent, diagnose, and treat disease" [13]. Moreover, the approach of PM is the category of medical sciences that aims to predict the possibility of developing a disease, achieve a precise diagnosis, and optimize the best performance of treatment for a particular patient [14]. In other words, according to figure 2, precision medicine intends to consider individual patient variables in terms of genetic, lifestyle, and environmental effects also distinguishing patients from other patients with the same presentations for improving the clinical practice, minimizing side effects, and increase the performance of treatment [15].

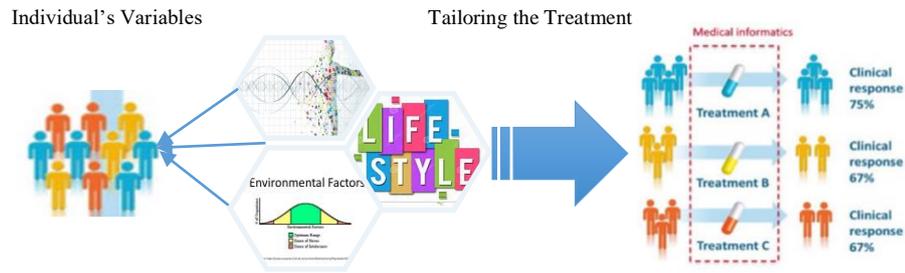

Figure 2. Precision Medicine

## 4. BARRIERS of PRECISION MEDICINE ADOPTION

### 4.1. Characteristics of the Domain

#### 4.1.1. Adoption of Classification Systems and Clinical Terminologies

Although terminologies that contribute to PM have been built using ontologies (a form of reasoning through logics for classifications), the challenges related to the employment and integration of them in healthcare is one of the major barriers that strongly affect the development of PM application [16], [17].

Lack of nation-wide projects and less completed projects of implementing ontology-based, terminologies have caused the stakeholders to avoid solid conclusions about the projection of ontology-based terminologies. Moreover, the lack of experts in biomedical ontologies and the semantic web are another limitation for PM development. As a result, the architectures to fully integrate different types of ontologies are still immature. This lack of large developments resultsin a high risk when designing EHR systems that use multiple ontologies since a large number of terminologies and ontologies have been developed in parallel by different bodies, thus making ontology mapping among them extremely challenging[18], [19].

#### 4.1.2. Limitations in Clinical Evidence, Outcomes, and Value Assessment Practice

Whereas, technologies are employed to develop the new approach of Medical Decision Making; PM, insufficient information, successful clinical evidence, and unavailability of universally adopted data models for validating the practical work have not guaranteed the value and outcome of PM [16].

#### 4.1.3. Adoption of Standards for Data Collection and Integration- Clinical Data Exchange

There have been various limitations and challenges in laboratory, medication, diagnosis, radiology, pathology, clinical evidence & outcomes, and procedures, based on the availability of different data structures and related challenges of adopting the existing standards for data exchange. For example, incorporation of genetic results into EMR in a searchable way. Moreover, in many cases, tests, which are conducted in external labs, cannot be integrated into other systems. Differences between lab code systems and identifications or missing the genetic info in EHRs data are some of the key challenges related to the data exchange [16]. Furthermore, identifying the right moment for clinical data exchange becomes a competitive process that needs attention [20].

The researchers have identified the need of extending the scope of the existing standards rather than inventing a new one where data standardization for integration and exchange is required for the correct interpretation of the data elements.

Although various initiatives have worked in this line to facilitate the adoption of the data standards especially in the omics discipline such as BioSharing, (works to ensure the standards are searchable and informative.), to extend the existing standards, the relevant stakeholders should cooperate to educate the potential adopters for understanding and using the existing data standards [21].

**4.1.4. Data Processing and Storage – Handling Big Data**

The speed of generating a massive amount of data from various resources has not been balanced with big data management yet. The analysis and processing of such data need hardware and software facilities, which is beyond the local infrastructures of many small laboratories. Particularly, molecular and omics data analysis requires powerful computational tools [21].Whereas componental resources and related facilities require a high cost of maintenance and development [6], adopting technologies such as cloud computing and the Internet of Things (IoT)can be an alternative to deal with the high cost of storage and computational requirements. For example, Siemens Healthcare, Philips Healthcare, and GE Healthcare are the three big medical imaging companies that pioneered switching computer-intensive image processing to the Health Cloud ecosystem. Together they have developed a strategy for establishing an "All-in-one Health Cloud" to provide the unique cloud platform [2], [22]. Although cloud computing can be a potential solution to deal with big data management issues, the need for protecting healthcare data and applying security and privacy levels will be increased[6], [23]. To deal with this requirement, solutions such as access control models and blockchain technology are offered. However, each solution carries specific limitations, which need to be addressed. For example, access control models, which perform based on the access authorizations and defined regulations, are less effective against internal risk attacks. Hence, in many cases, hybrid approaches are applied to integrated access control with some other methods such as encryption. Moreover, Blockchain technology is another alternative for secure data management in cloud platforms where the possibility to delete or modify stored data is almost zero. This characteristic of blockchain criticizes the regulations of personal data privacy; when it is obligated to delete the personal data based on individual patient's rights. Hence, in this case, blockchain technology requires to solve these types of challenges through new policies, privacy models, and further studies[24].

**4.1.5. Ethical and Legal Literature for Open Data Exchange**

Data protection, security, and related ethical issues are another challenge. Personal health data is sensible and valuable. It has been estimated that the healthcare domain is 200 % more likely to experience a data breach than other industries. Moreover, because data generated modular via the internet and in many cases stored offline, ethical consideration prevents an open data exchange [6]. For example, unintended access (the misuse of information gained through unauthorized access) is a serious challenge when it reaches to debating about ethical issues [25].

Due to the reforming of the medical domain as the data-sharing community, the patient's role as the owner and manager of her/his health data became more valid. Thus changing the power between care providers and care receiver is considerable needs restructuring regulations and policies. On one hand, the patient is requested to share personal information, on the other hand, it gives the right to the individual to become a member of the data-sharing community thus the roles, responsibilities, and authorities need restructuring and ethical issues should be addressed in an effective way [3].

### 4.1.6. Features of Healthcare Data

For PM, the major data include EHRs and Omic. Big Omic data contains a comprehensive catalog of molecular profiles (e.g., genomic, transcriptomic, epigenetic, proteomic, and metabolomics) and EHRs contain structured and unstructured clinical data. Both Omic and EHRs are challenging for analytics because of the data frequency, quality, dimensionality, and heterogeneity. Moreover, EHRs include structured data (e.g., ICD-9 diagnosis codes, administrative data, charts, and medication) and unstructured (e.g., clinical notes). Structured data includes two classes: administrate data, which remain unchanged during the entire clinical process. For example, demographics are administrative data. The second class is ancillary clinical data; the data, which frequently or continuously recorded during the clinical process such as medications and lab tests of blood pressure monitoring via sensors. This heterogeneity of data is a considerable challenge for data analysis[26]. Furthermore, in omic data, the combination of biological, instrumental, and environmental factors affects the quality of data. In EHRs, the quality of data depends on the missing data and incorrect records, which are the consequences of the clinician's teamwork to enter data, miss-interpretation, and the organization of documents for data entry. Hence, any mistakes and misinterpreting will cause wrong conclusions and inaccurate decision-making [26].

### 4.1.7. Adoption of Genomics in EHR (eMERGE)

Adoption of PM requires the integration of EHRs and Genomics (eMERGE). In a clinical setting, EMR (Electronic Medical Records) is used for clinical decision support and the integration between genomic/genetic data and EMR identifies causal genomic variants and genotype-phenotype associations into the EMR system. The major challenge of such integration is finding the most suitable method for storing and reprocessing the variants present in an individual or even family and the next generation in EMR. Moreover, for integrating genetic/genetic data into the EHRs, the size of genetic/genomic laboratory test results and the limited capacity of EHRs also has been identified as significant limitations. Because each individual has millions of variants and the variation of genomic data cannot fit into the current design of EHRs. One potential solution is to archive raw data in separate data repositories to be accessible once it is requiring. However, storing such unstructured data makes the processing speed slow. Another alternative is to strategize the external genetic/genomic data warehousing. In this approach, data is stored external to EHRs and the link connect and integrate particular record to related EHRs record [27].

In the addition, the genetic records stored in EHRs need to be interpreted throughout the clinical terms and vocabularies. To tackle this challenge, a couple of solutions have been suggested such as using rule-based decision support systems and using visualization elements for better presentation.

Besides, EHRs include medical records of all participants and hospitals, while EMR contains limited data for local clinics and hospitals. Therefore, standardizing the data exchange and defining protocols for better interoperability is useful for effective treatment [26]. Moreover, in addition to the technological challenges for adopting genomic/genetic in EHRs, the environmental aspects such as ethical and legal boundaries for data sharing and such integration need attention[17], [28]. To support the adoption of PM, various countries propose practical policies to protect individual's genomic/genetic data from discrimination to assure the patient about the security and privacy level of their data [29].

### 4.1.8. Research and Practice

Many research works have discussed the effective role of AI/ML for early diagnosis and better treatment, but also the relevant challenges have been addressed. Since projecting such approaches is still in the development stage, the lack of understanding of AI and ML is a limitation, and identifying the most effective approaches for pioneering valid clinical practice of PM needs more scientific research and practice [16]. Furthermore, the techniques for automating data collection, analysis, and processing are usually projected locally, and transferring the success story to other healthcare providers is not easy [6].

To speed up the delivery of precision medicine, more research is needed, particularly in the following biomedical big data areas and omics data integration. The big omics data analysis, for example, provides a holistic view for analyzing the patient condition and for effective prediction [26]. in many cases, applied research environments and academia have faced limitations and challenges to accessing healthcare data, in terms of privacy and security. Thus, Protected Health Information (PHI) and lack of trust are some of the critical issues that affect the development of research works in this area [16].

The second area in which research and practice demonstrate the influential effect is Patient segmentation based on similarities. Classifying patients based on biological similarities on their profile is a crucial phase in developing PM. Therefore, data mining applied in EHRs us clinical indicators such as drug responses, physiological signals, and disease susceptibility for patient classifications. However, high patient variability for disease and also the fact that many subgroups of the disease have not been identified yet, affect the performance of the practical PM application. Hence, this gap needs systematic research for validating the classifications of the patient based on EHR mining [26].

Finally, as was discussed above, healthcare requires strong clinical evidence to analyze the validity of precision medicine in practice. This requirement is completely important for expanded clinical use [5]. Therefore, to address the limitations, it is required to develop secure research-based frameworks for efficient data collection, data integration, storage and pre-processing, de-identification to serve a large community of users, support organizational policies, and provide efficient access and connectivity [16]. Such as a new research platform that minimizes the isolation between scholar platform and clinical data [21].

### 4.2. Lack of a Unique Definition of Precision Medicine

Although Precision Medicine is in use as a common label, by many public funding streams (e.g., Genome England, Australian Genomics, and the Center for Personalized Cancer Treatment in the Netherlands) and private funding streams (e.g., IBM Watson), but the terminology is still evolving, and multiple terms have been used interchangeably literature has used various terms such as 'pharmacogenomics' and 'P4 medicine' (preventative, predictive, participatory, and personalized), to point out different aspects of the relevant research about precision medicine. Whereas all of these conceptualizations seem to perform in the same direction [3], the lack of the unique terminology may lead to misunderstanding by the stakeholders who have the power and responsibilities and different roles in accelerating PM adoption [5], [30].

## 5. EXAMINING THEORIES; DISCUSSION and FINDINGS

From the socio-tech theory perspective, in the implementation of PM application two subgroups play a critical role: technological and social. Also, the effective balance between the two aspects grantees success. Moreover, as we discussed above, the diffusion process requires, innovation to be communicated in time through social channels. Combining the major success indicators of these two theories for examining PM application implementation, we conclude that the PM application as the innovation/technology needs to be assessed in terms of the degree of its compatibility with existing systems in healthcare (e.g., workflows, business processes, hardware/software infrastructures and resources). Moreover, the benefit and advantages of using this technology should be identified and communicated widely via a social platform, where the power and role of stakeholders' networks are remarkably important. Besides, identifying the complexity, reliability, and observability of this technology needs to be evaluated via a defined research platform where the effective cooperation between academia, the healthcare industry, and technology providers results in trust and confidence to achieve the outcome of technological evaluations.

From the social and organizational point of view, as it was discussed in both theories, management support, effective communication, educating people, pipelining strategic plans for tackling the environmental pressures such as boundaries and policies make the diffusion smooth clear, and rewarding.

The adoption of precision medicine not only will affect many health systems, but also all stakeholders. (e.g., Individual patients, care providers, policymakers, technology providers, drug developers). Therefore, each level of stakeholders carries specific preferences, definitions, and the requirement for accepting the new innovative technology (PM); proposing the value assessment of the PM can be an alternative to link stockholders and facilitate and strategies education and communication required for new technological adoption.

Although the socio-tech and DoI, both address the social, organizational, and technological indicators in total, these two theories carry gaps. The limitations are identified where the characteristics of the domain and the vital needs for research and practice are discussed. As table 1, demonstrates, without solving the problems carries by healthcare (e.g., projection of terminologies and classifications, integrating genomes in EHRs, adopting standards for data exchange, challenges about data features, data security, and ethical issues, dealing with big data, selecting the most effective approaches), the emerging of PM will not be projected successfully. Moreover, presenting evidence-based validity, value assessment, proposing the unique and fashionable term and definition for PM, and identifying the most effective techniques and tools for data analysis, data extraction, and mining to fulfill the PM approach requires research and practice. This cooperation needs inputs from other disciplines such as biomedical research, statistics, economics, and ethics, and key health stakeholders; where, in practice, the interplay between several stakeholders should be taken into the account[5], [30], and finally, empowering patients, doctors and public, as the co-researcher about PM accelerate the adoption process. Making PM comprehensible, trustworthy, and equally accessible for the population as the participatory opportunities for people to participate in the production of data for the good of medical progress. Where the new forms of participation and data control are needed to be considered[31].

Table 1. Major considerations and limitations of PM adoption

| Major Limitations and Considerations of PM Adoption in Healthcare | Diffusion of Innovation | Socio-technical | Characteristics of Domain | Research &Practice |
|---|---|---|---|---|
| Adoption of classification systems and clinical terminologies | | | X | X |
| Limitations in clinical evidence & outcomes | X | | | X |
| Adoption of standards clinical data exchange | | | X | |
| Data processing and storage – handling big data | | | X | |
| High security, ethical and legal literature | X | | X | |
| Features of healthcare data(format, quality) | | | X | |
| Adoption of genomics in EHRs (emerge) | | | X | |
| Selecting the most effective approaches-techniques | | | X | X |
| Patient segmentation based on similarities | | | | X |
| Lack of a unique definition of PM | X | | | X |
| Management aspects<br>Education and skills<br>Communication<br>Environmental pressures (policies, regulations) | X | X | | X |

As a result, based on identified limitations of both theories (DoI, socio-tech) in IS implementation, figure 3 presents the final influential indicators for emerging PM adoption in healthcare. Where the two new aspects: "research & practice" and "characteristics of the domain" among the technology/innovation aspects and social/organizational elements are introduced.

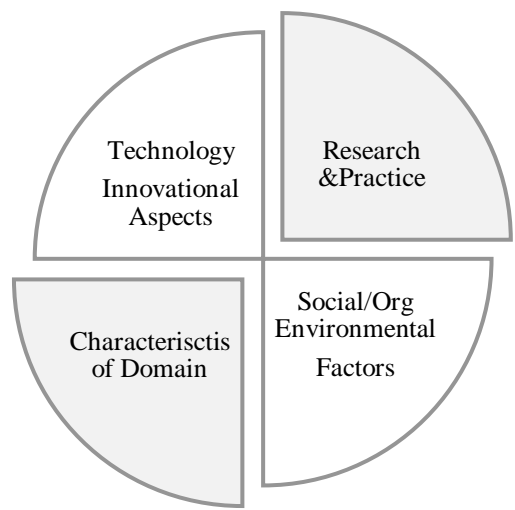

Figure 3. Success Indicators for PM Implementation- Theory Assessment

In summary, it can be concluded that PM is applicable to be introduced as socio-tech /DoI phenomena. Therefore, for integrating PM in healthcare, not only managerial and organizational aspects needs to be addressed, but also the technical issues such as the unique definition of PM, reliability, and complexity needs considerations.

Besides, there are major limitations, which are directly related to the features of the healthcare ecosystem itself. As is mentioned above, since healthcare is a data-driven platform, therefore, limitations and challenges carry by the characteristic of data (eg., sensitive, fragmented, heterogeneous, unstructured, high volume) strongly influence this transition. For example, we discussed the possibility of employing the cloud computing solution to deal with big data management in terms of storage management and maintenance, it is also discussed the limitations that cloud computing presents in terms of security and data protection. Moreover, where blockchain technology can deal with security and data protection, the role of individuals as the owner of data and their willingness for data sharing, open another limitation. Besides, not only data characteristics are a matter but also the complexity of the domain with interconnected entities provide complicated regulations that need attention. For example, projecting PM in healthcare needs a trustful cooperation platform for data sharing by empowering individuals to participate in their health management purpose. However, such an arrangement demands a deep restructuring of standards and policies about security and data protection.

## 6. CONCLUSIONS

This paper discussed the key limitations and considerations for emerging Precision Medicine in healthcare. Although there have been various terms used interchangeably as PM, it is introduced as a new approach in medical decision making where the individual patient variables in genetics, lifestyle, and environmental are taken into the account. This study employed the socio-tech and DoI paradigms to explain the various technological (e.g., complexity, compatibility, trialability, tangibility, perceived value), social-organizational (communication, education, cooperation, management), and environmental (e.g., policies, regulations, ethical issues) indicators which influence the successful implementation. By examining the combined theories (socio-tech, DoI), the PM application is presented as a socio-tech and DoI phenomena through this theoreticalassessment, we identified two critical aspects, which are missed to address by both theories: "characteristics of healthcare" and "research and practice under a cooperative model of partnership and trust". Accounting for these two identified aspects enables us to deal with the limitations and difficulties we have faced in a complex domain like healthcare. Besides, to facilitate the transition through valid clinical evidence and value assessed outcomes of PM practice. Successful adoption of PM in healthcare contributes to maximizing the quality of treatment and saving more lives by minimizing medical errors, decreasing the risk of over-treatment, minimizing the side-effect of medicines, and identifying the patient at risk for a particular disease.

## ACKNOWLEDGMENTS

The work has been supported by FCT – Fundação para a Ciência e Tecnologia within the Projects Scope: DSAIPA/DS/0084/2018.

**AUTHORS**

**Manuel Filipe Santos** received his Ph.D. in Computer Science (Artificial Intelligence) from the University of Minho (UMinho), Portugal, in 2000. He is an associate professor with habilitation at the Department of Information Systems, UMinho, teaching undergraduate and graduate classes of Business Intelligence and Decision Support Systems. He is the head of Intelligent Data Systems lab and the coordinator of the Information Systems and Technology group (www.algoritmi.uminho.pt) of the R&D ALGORITMI Centre, with the current research interests: Business Intelligence; Intelligent Decision Support Systems; Data Mining and Machine Learning (Learning Classifier Systems); and Grid Data Mining. He is part of the steering committees of the master's course in Engineering and Management of Information Systems and the Doctoral Program in Information Systems and Technology.

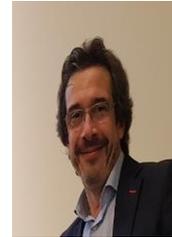

**Nasim Sadat Mosavi** is a Ph.D. student at the University of Minho (UMinho), Portugal. She also works as a researcher at Centro algoritmi, (UMinho). Her research interest is Intelligent Decision Support Systems (IDSSs) using Machine learning and optimization techniques. Healthcare/Medicine is her research interest domain. Nasim graduated in Computer Science (associate's degree) and Computer Engineering-software (bachelor's degree) from the Islamic Azad University of Tehran-Iran and she pursued her master's degree in International Business from the University of Wollongong. She was involved, in the development and implementation of more than 100 successful IS projects in different positions with different industries in Dubai, Iran, K.SA.

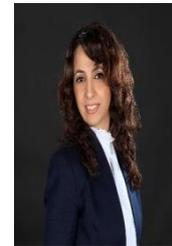